\documentclass[a4paper,twocolumn,floatfix,10pt,superscriptaddress]
{quantumarticle}
\pdfoutput=1

\usepackage[english]{babel}
\usepackage{amsmath}
\usepackage{bbm}
\usepackage{verbatim}
\usepackage{graphicx} 
\usepackage{amssymb}
\usepackage[colorlinks=true,pdfencoding=auto]{hyperref}
\usepackage[numbers,sort&compress]{natbib}
\usepackage{xcolor}
\usepackage{orcidlink}

\definecolor{mcrcolor}{rgb}{0.7,0.2,0.7}
\definecolor{aprcolor}{rgb}{0.7,0.1,0.2}
\definecolor{philcolor}{rgb}{0.2,0.6,0.2}
\definecolor{dptcolor}{rgb}{0.1,0.1,0.8}

\newcommand{\msf}[1]{\mathsf{#1}}
\newcommand{\mcl}[1]{\mathcal{#1}}
\newcommand{\bsb}[1]{\boldsymbol{#1}}

\newcommand{\bomega}{{\bar{\omega}}}

\newcommand{\mcH}{{\mcl{H}}}
\newcommand{\mcL}{{\mcl{L}}}

\newcommand{\msH}{{\msf{H}}}
\newcommand{\msG}{{\msf{G}}}

\newcommand{\bu}{{\bsb{u}}}
\newcommand{\bv}{{\bsb{v}}}

\newcommand{\rmout}{{{\rm out}}}
\newcommand{\rmin}{{{\rm in}}}

\begin{document}

\title{Floquet Topological Frequency-Converting Amplifier}

\author{Adrian Parra-Rodriguez\,\orcidlink{0000-0002-0896-9452}}
\email{adrian.parra.rodriguez@gmail.com}
\affiliation{Instituto de F\'isica Fundamental (IFF), CSIC, Calle Serrano 113b, 28006 Madrid, Spain.}
\affiliation{Technical University of Munich, TUM School of Natural Sciences, Physics Department, 85748 Garching, Germany}
\affiliation{Walther-Meißner-Institut, Bayerische Akademie der Wissenschaften, 85748 Garching, Germany}
\affiliation{Munich Center for Quantum Science and Technology (MCQST), 80799 Munich, Germany}
\author{Miguel Clavero-Rubio\,\orcidlink{0009-0002-7132-7097}}
\affiliation{Instituto de F\'isica Fundamental (IFF), CSIC, Calle Serrano 113b, 28006 Madrid, Spain.}
\author{Philippe Gigon\,\orcidlink{0009-0008-9802-9261}}
\affiliation{Technical University of Munich, TUM School of Natural Sciences, Physics Department, 85748 Garching, Germany}
\affiliation{Walther-Meißner-Institut, Bayerische Akademie der Wissenschaften, 85748 Garching, Germany}
\affiliation{Munich Center for Quantum Science and Technology (MCQST), 80799 Munich, Germany}
\author{Tom\'as Ramos\,\orcidlink{0000-0003-2182-7878}}
\affiliation{Instituto de F\'isica Fundamental (IFF), CSIC, Calle Serrano 113b, 28006 Madrid, Spain.}
\author{Álvaro Gómez-León\,\orcidlink{0000-0002-3990-5259}}
\affiliation{Instituto de F\'isica Fundamental (IFF), CSIC, Calle Serrano 113b, 28006 Madrid, Spain.}
\author{Diego Porras\,\orcidlink{0000-0003-2995-0299}}
\email{diego.porras@csic.es}
\affiliation{Instituto de F\'isica Fundamental (IFF), CSIC, Calle Serrano 113b, 28006 Madrid, Spain.}

\begin{abstract}
We introduce a driven–dissipative Floquet model in which a single harmonic oscillator, with both frequency and decay rate modulated, realizes a non-Hermitian synthetic lattice with an effective electric-field gradient in frequency space. Using the Floquet-Green's function and the doubled Hamiltonian representation of non-Hermitian matrices, we show that the linear response of this system is characterized by a local winding number. Nontrivial values of the winding number induce directional amplification in the synthetic dimension, thereby converting input signals to different frequencies. The underlying mode structure is well described by a Jackiw–Rebbi-like continuum theory with Dirac cones and solitonic topological zero modes in synthetic frequency. Our results establish a simple and experimentally feasible route to non-Hermitian topological amplification, naturally implementable in current quantum technologies such as superconducting circuits.
\end{abstract}

\maketitle

\section{Introduction}
Topology has transformed our understanding of quantum systems by revealing new classes of states characterized not by local order parameters but by robust global topological invariants. Initially developed in the study of electronic systems, such as quantum Hall phases and topological insulators \cite{Hasan2010,Qi2011}, these ideas have spread across a broad range of physical settings. Topological photonics ~\cite{Lu2014,Ozawa2019,Mehrabad23} extends these concepts to bosonic modes of light, enabling applications like robust control over signal propagation ~\cite{Barik2018} and topological frequency combs~\cite{Flower2024}. In these systems, the combination of time-reversal symmetry breaking and non-Hermiticity due to intrinsic gain and loss leads to nonreciprocal transport and amplification~\cite{Metelmann15,Fang2017,peano2016,McDonald2022, delPino2022,Busnaina2024, Slim2024,Ramos2022}. A general theoretical framework based on topological band theory (TBT) extended to non-Hermitian systems \cite{Esaki2011,Gong2018} can be used to describe topological amplification in photonic lattices \cite{Porras2019,wanjura2020topological}.
In this approach, topological features emerge in the singular value spectrum of the coupling matrix in input-output theory, leading to directional amplification that remains robust in the presence of disorder 
\cite{Ramos2021, Wanjura2021,Gomez2022, GomezLeon2023, ClaveroRubio2025}. 

Floquet systems, in which one or more parameters are periodically driven in time, further expand the landscape of topological phases. A prominent example is the class of Floquet topological insulators: nonequilibrium states of matter in which periodic driving induces novel topological phases that may be absent in equilibrium~\cite{Lindner2011,Rechtsman2013,GomezLeon2013,Baum2018}. Such periodically driven systems are characterized by the emergence of one or more additional effective dimensions in frequency space. It has been shown that multiple incommensurate drives can simulate topological band structures in those synthetic dimensions, leading to topological frequency conversion~\cite{Martin2017} or quantized energy pumping between driving sources~\cite{Luneau2022}. Furthermore, periodic drivings have been used to improve traveling-wave amplifiers~\cite{Peng2022}.

Latest advances have highlighted complementary mechanisms for engineering topology in driven photonic platforms. In particular, time-Floquet modulation can induce nonreciprocal gain and directional amplification in zero-dimensional parametric systems~\cite{Koutserimpas2018}, static dissipation can be used to autonomously stabilize nonequilibrium Floquet states~\cite{Ritter2024}, and periodic modulation of a single microwave resonator has already been used to study synthetic Bloch-wave dynamics \cite{Ahrens2025}. 

In this article, we demonstrate that periodically driven dissipation in addition to frequency modulation is sufficient to generate stable topological amplification and directional frequency conversion in an extremely minimal setting: a single bosonic mode subjected to a single periodic modulation of both its frequency and decay rate. The resulting driven–dissipative dynamics define a non-Hermitian lattice with an effective electric-field gradient in frequency (Floquet–Sambe) space, locally analogous to a Floquet version of the Hatano–Nelson (HN) model \cite{Hatano1996} featuring asymmetric synthetic hopping between neighboring harmonics. A useful qualitative picture is provided by the Jackiw–Rebbi (JR) model \cite{Jackiw1976}, a one-dimensional Dirac equation with a sign-changing mass that hosts a topological zero mode at a domain wall, a minimal situation where index theorems apply \cite{AtiyahSinger1963,Callias1978}. The topological properties of our effective Floquet lattice are revealed through the definition of a local topological invariant related to the system’s Green’s function, which determines amplification bandwidth, directionality, and robustness \cite{yao2017topological,Yao2018}.

Although our model resembles certain features of these earlier approaches, it operates through a fundamentally different physical mechanism. Here, topology is generated via competing coherent and dissipative Floquet couplings on a single-mode platform, thereby avoiding the need for multifrequency drives \cite{Martin2017,Crowley2019,Koch2024} or multimode architectures \cite{Koutserimpas2018,KochF2025}. Following well-established topological input–output theory~\cite{Ramos2021}, we extract the associated local topological invariant, compute a signal-to-noise ratio (SNR), and validate our predictions through exact numerical simulations. Finally, we put forward a feasible experimental implementation based on superconducting circuits, in which the required driven dissipation emerges naturally through the adiabatic elimination of rapidly decaying ancillary modes, thereby enabling stable intraband microwave-to-microwave frequency conversion.
\begin{figure}[t]
\centering
\includegraphics[width=1\linewidth]{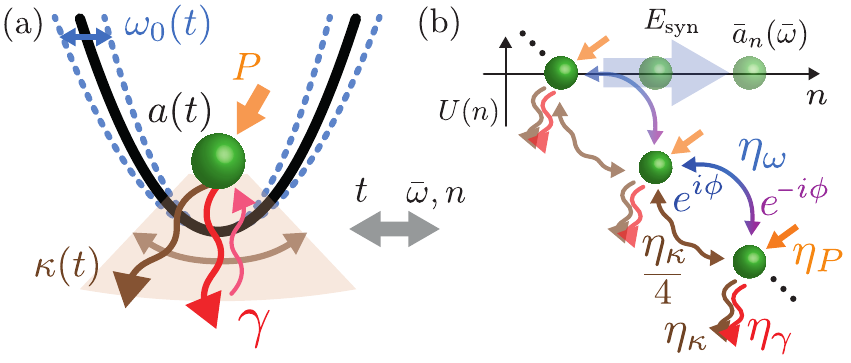}
\caption{Sketch of the effective model: (a) A confined particle in a periodically driven harmonic potential $\omega_0(t)$, coupled to a static decay channel $\gamma$ (also serving as a coherent drive port), a time-modulated decay channel $\kappa(t)$, and an incoherent pump $P$. (b) These periodic modulations implement a synthetic lattice with an effective constant electric field $E_{\rm syn}$ in frequency space, giving rise to directional amplification and frequency conversion due to time-reversal-symmetry breaking via asymmetric effective couplings.}
\label{fig:Model_Intro}
\end{figure}

\section{Model} We consider an open quantum system consisting of a single bosonic mode $a$, whose frequency $\tilde{\omega}_0(t)=\omega_a+\omega_0(t)$ and decay rate $\kappa(t)$ are periodically modulated, together with a static decay channel $\gamma$, and an incoherent pump $P$; see Fig.~\ref{fig:Model_Intro}(a). In the rotating frame with respect to the bare cavity frequency $\omega_a$, the dynamics are governed by the master equation 
\begin{align}
\mcL(\rho) = -i[H(t), \rho] + (\gamma+\kappa(t))\, \mcl{D}[a]+ P\, \mcl{D}[a^\dag],
\end{align}
with the time dependent Hamiltonian 
\begin{align}
    H(t)=
    \omega_0(t)a^\dagger a
\end{align}
and the coherent periodic modulations 
\begin{align}
    \omega_0(t) &= 2 \eta_\omega \Omega \cos(\Omega t + \phi),\\\kappa(t)&=2\eta_\kappa \Omega \cos^2(\Omega t/2).
\end{align} Here, $\Omega$ denotes the common modulation frequency, whereas $\eta_\omega,$ and $\eta_\kappa$ are the modulation amplitudes for frequency and decay, respectively. For convenience, we also express the incoherent pump and the static decay gamma in terms of the modulation amplitude as $P=\eta_P \Omega$, and $\gamma=\eta_\gamma\Omega$.
 Without loss of generality, we assume throughout that $\eta_\alpha>0$ for all $\alpha\in \{\omega, P, \kappa, \gamma\}$.

The dissipative dynamics takes into account the effects of a time-dependent decay rate, a constant incoherent pump, and a static loss channel, via a Lindblad term of the form $\mcl{D}[A] = A \rho A^\dag - \tfrac{1}{2} A^\dag A \rho - \tfrac{1}{2} \rho A^\dag A$. As we show in Sec. \ref{Sec::IMP}, the modulated decay can be realized via a time-dependent coupling $g_b(t)$ to a fast, strongly damped auxiliary mode $b$ with decay rate $\kappa_b \gg \gamma, g_b(t)$, while the incoherent pump can be implemented through coupling to a fast-decaying auxiliary mode $c$ ($\kappa_c \gg \gamma, g_c$) via a parametric interaction (see Appendix \ref{app:AElimination} for details on the adiabatic elimination).
 
\section{Input--output theory} 
\label{sec:I-O_theory}To analyze this dynamics, we employ quantum Langevin theory \cite{Ramos2021}. In the rotating frame, the single mode obeys
\begin{align}
\dot{a}(t) = -i \omega_0(t)\, a(t) 
+ \frac{P -(\gamma+ \kappa(t))}{2}\, a(t)
+ \xi_\rmin(t), \label{eq:adot}
\end{align}
with the composite input operator
\begin{align}
    \xi_\rmin(t) = - \sqrt{\gamma}\, d_\rmin(t)- \sqrt{\kappa(t)}\, w_\rmin(t) +\sqrt{P}\, h_\rmin^\dagger(t),
\end{align} where $d_\rmin$, $w_\rmin$ and $h_\rmin$ are input noise operators for the $\gamma$, $\kappa(t)$, and $P$ independent channels, respectively.

Because the dynamics is linear and periodic, the system admits a Floquet pseudo--steady state. In this regime, the Fourier--Floquet expansion is well defined,
\begin{align}
a(t) = \frac{1}{\sqrt{2\pi}}\int_0^{\Omega} d\bomega\; \sum_{n} \bar{a}_n(\bomega) \,
e^{-i(\bomega + n \Omega) t},\label{eq:at_a_n_bomega}
\end{align}
with $\bomega\in[0,\Omega)$, and analogously for the noise operators. In this basis, the equations of motion reduce to a linear algebra problem with formal solution \begin{align}
    \bar{a}_n(\bomega) 
= i \sum_m \msf{G}_{nm}(\bomega)\, \bar{\xi}_{\rmin,m}(\bomega),
\end{align} 
where the Floquet--Green's function (FGF) is given by
\begin{eqnarray}
    \msf{G}(\bomega) \equiv \big(\bomega - \bar{\mathsf H}\big)^{-1}.
\end{eqnarray}
The effective dynamical matrix in frequency space reads
\begin{align}\label{eq:Hnm_Sambe}
    &\frac{\bar{\mathsf{H}}_{n m}}{\Omega} = \left[ - n
+ i\eta_\kappa\left( \frac{\beta-1}{2} \right) \right] \delta_{n,m} \\
&+\left( \eta_\omega e^{i\phi} - \frac{i \eta_\kappa}{4} \right)\delta_{n,m-1}
+ \left( \eta_\omega e^{-i\phi} - \frac{i \eta_\kappa}{4}\right)\delta_{n,m+1}.\nonumber
\end{align}
where we have defined the static-dynamic dissipation ratio (analogous to previously defined quantity in Ref.~\cite{wanjura2020topological}) as $\beta\equiv(\eta_P-\eta_\gamma)/\eta_\kappa$. This parameter physically quantifies the effective competition between incoherent pump and static losses, normalized by the strength of the modulated dissipation that induces asymmetric synthetic hopping. By direct inspection of Eq.~(\ref{eq:Hnm_Sambe}), one can verify that $\beta$ governs the system’s stability by controlling the sign of the complex term in the diagonal. Further discussion about stability is provided in Sec. \ref{sec:Stability_SNR}.

This structure arises because both $\omega_0(t)$ and $\kappa(t)$ contain harmonics at frequency $\Omega$, leading to nearest-neighbor couplings in the synthetic lattice. This formulation, which locally resembles a Floquet generalization of the HN model, captures directional frequency conversion and amplification. The linear $n$-dependence of the diagonal term can be interpreted as a potential $U(n)=-n$ induced by a synthetic electric field, $|E_{\rm syn}| = 1$ \cite{GomezLeon2013,Martin2017}.


The input noise vector in frequency space is given by
\begin{align}\label{eq:noise_op_bomega}
\begin{aligned}
\bar{\xi}_{\rmin,n}&(\bomega)/\sqrt{\Omega}
= \,\sqrt{\eta_P}\, \tilde{h}^\dagger_{\rmin,-n}(-\bomega)- 
\sqrt{\eta_{\gamma}}\, \bar{d}_{\rmin,n}(\bomega)\\
&- \sqrt{\frac{\eta_\kappa}{2}} \left( \bar{w}_{\rmin,n}(\bomega + \frac{\Omega}{2}) + \bar{w}_{\rmin,n-1}(\bomega + \frac{\Omega}{2}) \right),
\end{aligned} 
\end{align}
for $\bomega \leq \Omega/2$. If 
$\bomega > \Omega/2$, the last term must be replaced by $-\sqrt{\eta_\kappa/2}
\left( \bar{w}_{\rmin,n+1}(\bomega - \Omega/2) + \bar{w}_{\rmin,n}(\bomega - \Omega/2) \right)$. Note that the last term introduces $\Omega/2$ harmonics through the modulation of the coupling to the auxiliary cavity, but, assuming the latter is in vacuum, these harmonics do not contribute\footnote{In a more general situation, e.g., if the auxiliary cavity were thermally occupied or driven, these $\Omega/2$ sidebands would matter. Such cases can be conveniently treated by sampling the Floquet-Sambe space at spacing $\Omega/2$, avoiding the split definition of the noise operators in Eq.~\eqref{eq:noise_op_bomega}.}. 

Using the input–output relation for the statically-coupled dissipative port $\gamma$, \begin{align}
    d_{\rmout}(t)=d_{\rmin}(t)+\sqrt{\gamma} a(t),
\end{align} together with the linear solution above, we can express the output field in frequency space as 
\begin{align}
    \bar{d}_{\rmout,n}(\bomega) 
= \bar{d}_{\rmin,n}(\bomega) 
+ i \sqrt{\gamma} \sum_m \msf{G}_{n m}(\bomega)\, \bar{\xi}_{\rmin,m}(\bomega).\nonumber
\end{align}
Substituting the input vector $\bar{\xi}_{\rmin,m}(\bomega)$, we obtain
\begin{align}
\bar{d}_{\rmout,n}(\bomega) 
&= \sum_m \msf{R}_{n m}(\bomega)\, \bar{d}_{\rmin,m}(\bomega)\label{eq:bar_dn_out}
\\&+\msf{P}_{nm}(\bomega) \bar{h}_{\rmin,-m}^\dag(-\bomega)+ \msf{Q}_{n m}(\bomega)\, \bar{w}_{\rmin,m}(\bomega),\nonumber
\end{align}
with effective scattering matrices
\begin{align}\label{eqs:R_nm_P_nm}
\begin{aligned}
    \msf{R}_{n m}(\bar\omega) &= \delta_{n m} - i \Omega\,\eta_{\gamma}\,\msG_{n m}(\bar\omega),\\
\msf{P}_{n m}(\bar\omega) &= i\,\Omega\sqrt{\eta_{P}\eta_{\gamma}}\,\msf{G}_{n m}(\bar\omega).
\end{aligned}
\end{align}
$\msf{Q}_{n m}(\bomega)$ involves the $\Omega/2$ sideband structure already mentioned and is omitted since that port is taken to be in vacuum.

These matrices contain all the necessary information to  assess the device performance in Sec.~\ref{sec:Stability_SNR}. Given a coherent drive applied at port $\gamma$, the corresponding input field is $\langle d_{\rm in}(t) \rangle = \alpha_d\, e^{-i\omega_d t}$ with $\omega_d = \bar{\omega}_d + n_d \Omega$, while all other inputs remain in vacuum. Using Eq.~\eqref{eq:bar_dn_out}, the mean output field in the Floquet--Sambe basis is given by 
\begin{eqnarray}
    \langle \bar{d}_{{\rm out}, n}(\bar{\omega}) \rangle = \sqrt{2\pi}\, \alpha_d\, \mathsf{R}_{n,n_d}(\bar{\omega})\, \delta(\bar{\omega}-\bar{\omega}_d).
\end{eqnarray}

\section{Topological frequency conversion and amplification}
The non-Hermitian Green's function of linear systems such as ours is well known to be linked to an eigenvalue problem in a doubled space~\cite{Porras2019,Ramos2021,ClaveroRubio2025,Gomez2022,GomezLeon2023,Herviou2019}. Explicitly,  the doubled Hermitian matrix
\begin{align}\label{eq:doubled_matrix}
\mcH(\bomega) =
\begin{pmatrix}
0 & \bomega - \tilde{\msH} \\
\bomega - \tilde{\msH}^\dag & 0
\end{pmatrix},
\end{align} satisfies the eigensystem 
\begin{align}\label{eigensystem}
\mcH(\bomega)\begin{pmatrix}
        \bu_l\\ \pm \bv_l
    \end{pmatrix}=\pm E_l \begin{pmatrix}
        \bu_l\\ \pm \bv_l
    \end{pmatrix}.
\end{align}
Because of Eq. \ref{eigensystem}, the singular value decomposition (SVD) of the Green’s function can be expressed in terms of the singular vectors and eigenfrequencies $E_l>0$, as
\begin{eqnarray}
    \msf{G}(\bomega) = \sum_l \bv_l\, \frac{1}{E_l}\, \bu_l^\dag.
\end{eqnarray}
 This correspondence establishes a link between input-output theory and TBT~\cite{Ramos2021}, thereby enabling the characterization of topological phases. In our case, $\mcH(\bar{\omega})$ possesses chiral symmetry (${\cal S}\mcH(\bar{\omega}){\cal S} = -\mcH(\bar{\omega})$, with ${\cal S} = \mathbb{I}\otimes\sigma_z$) by construction, and, in general, belongs to the AIII symmetry class of the Altland–Zirnbauer classification~\cite{Altland1997,Roy2017_FTI-classification}, supporting directional frequency conversion and amplification via topologically protected edge states.

\begin{figure}[t]
\centering
\includegraphics[width=1.01
\linewidth]{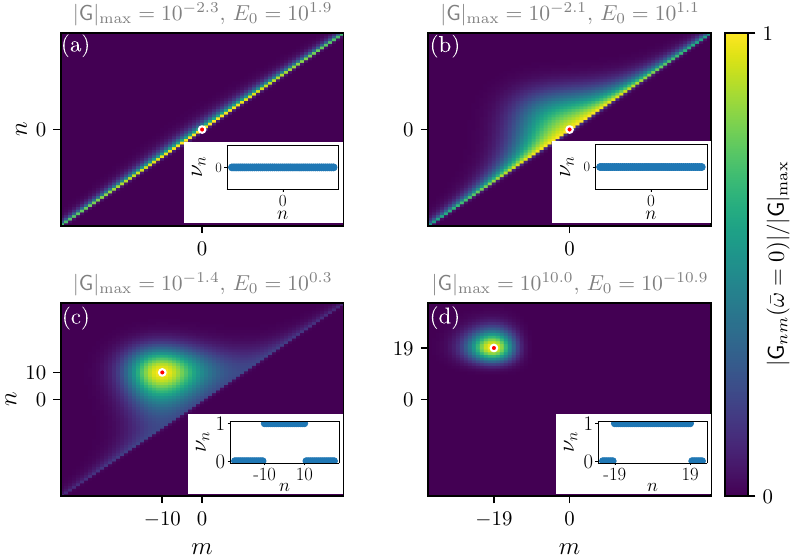}
\hspace*{-.5cm}\includegraphics[width=.88
\linewidth]{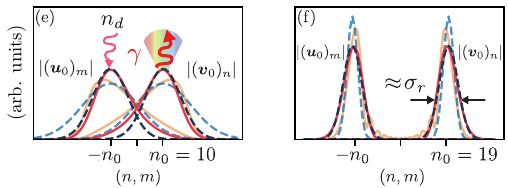}
\caption{(a–d) The local winding number $\nu_n(\bomega)$ accurately predicts the topological region where $|\msf{G}_{nm}(\bomega)|$ reaches its off-diagonal maximum, signaling strong frequency conversion and, as $\beta\to1$, enhanced amplification ($E_0\to0$). Parameters from (a-d) are $\eta_P = s \times (1,9.5,11.4,19.5)$ with $\eta_\omega = \eta_\kappa/s = \eta_\gamma/s = 10$, $\phi=\pi/2$, $\bomega=0$, $s=3$, and $\Omega = 2\pi$, leading to $\beta=(-0.9,-0.05,0.14,0.95)$.
(e,f) Normalized singular vectors for the parameter sets of panels (c) and (d), corresponding to $\beta = 0.14$ and $\beta = 0.95$ (topological stable regime $0<\beta<1$). Curves for scaling parameters $s=1$ (orange, blue) and $s=3$ (red, black) are shown. Solid lines come from numerical diagonalization of Eq.~\eqref{eq:Hnm_Sambe}, while dashed lines indicate the JR prediction (whose accuracy improves as $s\to\infty$).
(e) For an input signal at harmonic $n_d$ from the $\gamma$-port, weighted by $(\boldsymbol{u}_0)_{n_d}$, the system’s steady state (and emitted spectrum $\bar{d}_{\rm out,n}(\bomega)$) contains harmonics given by $(\boldsymbol{v}_0)_n$, centered at $n_0$ (with width $\sigma_{\rm r}\to\sqrt{\eta_\kappa/2}$ as $\beta\to1$), achieving maximum conversion when $n_d = -n_0$.}\label{fig:Gn_winding}
\end{figure}

Since the full model is infinite in the harmonic index $n \in \mathbb{Z}$, we truncate to a finite synthetic frequency window $|n| \leq N$, justified by the spectral tilt $-n\delta_{n,m}$ in Eq.~(\ref{eq:Hnm_Sambe}) acting as a synthetic electric field that detunes distant sites. We denote the truncated matrix $\tilde{\msH}$ and impose periodic boundary conditions to define the discrete Fourier transform $\bar{a}_k = \frac{1}{\sqrt{2N+1}} \sum_{n=-N}^N e^{-i k n} \bar{a}_n$, 
where $k = 2\pi m/(2N+1)$ with $m \in [-N,N]$. In this basis, the dynamical matrix becomes
\begin{align}\label{eq:H_kkp_final}
\begin{aligned}
\frac{\tilde{\msH}_{k k'}}{\Omega} &= 
\left( 2 \eta_\omega \cos(k + \phi) 
- i \frac{\eta_\kappa}{2} \cos(k) \right) \delta_{kk'} \\
&\quad + i\eta_\kappa\left( \frac{\beta-1}{2} \right)\delta_{kk'} - f_{kk'},
\end{aligned}
\end{align}
with $f_{kk'} = \frac{1}{2N+1} \sum_n n e^{-i(k - k')n}$ representing the tilt. Note that replacing $f_{kk'}$ by a $k$-independent constant restores exact translation symmetry and reduces Eq.~\eqref{eq:H_kkp_final} to the standard Hatano–Nelson model, whose phase diagram is well known. In the spirit of local topological markers~\cite{BiancoResta2011}, we use a local (frozen-$n$) approximation to retain the linear tilt while keeping a band picture, i.e., for each harmonic $n$ we set $f_{kk'} \to n\,\delta_{kk'}$ and analyze the resulting family of local Bloch Hamiltonians $\tilde{\msH}_{k}(n)$. Within each slice $(\bomega,n)$, we then define a local winding number
\begin{align}
\label{eq:winding_number_Hbar}
\nu_n(\bomega)
=\oint \frac{dk}{4\pi i}\, \mathrm{Tr}\!\left[\sigma_z\, \tilde{\mcH}_k^{-1}\partial_k \tilde{\mcH}_k\right],
\end{align}
following the standard definition of the Floquet winding invariant for chiral 1D systems~\cite{Kitagawa2010}. In the doubled representation, we express $\tilde{\mcH}_k(n)=\bsb{r}(k,n)\cdot\bsb{\sigma}$, with $\bsb{\sigma}=(\sigma_x,\sigma_y,\sigma_z)$, and $\bsb{r}(k,n)$ encoding the Bloch vector. 

Working at the maximally nonreciprocal phase, i.e., $\phi=\pi/2$, we can write the associated doubled-space dynamical matrix 
\begin{align}
\tilde{\mcH}_k(\bomega,n)=r_x(k,n)\sigma_x+r_y(k,n)\sigma_y \label{eq:H_k_omegabar_n}
\end{align}
with
\begin{align}\begin{aligned}
r_x(k,n)&=\bar{\omega} +\Omega\left(n+2\eta_\omega \sin k\right), \\
r_y(k,n)&=\tfrac{\Omega\eta_\kappa}{2}(\beta-1-\cos k).
\end{aligned}
\label{eq:rxry}
\end{align}
The associated winding number \eqref{eq:winding_number_Hbar}, equivalent to 
\begin{align}
    \nu_n(\bomega)=\oint \frac{dk}{2\pi}\,\frac{\big(\bsb{r}\times \partial_k \bsb{r}\big)_z}{\|\bsb{r}\|^2},
\end{align}
is $0$ or $\pm 1$ depending on whether the trajectory of $(r_x,r_y)$ encloses the origin. Furthermore, the sign of the winding number is sensitive to the curve orientation. In particular, $\mathrm{sgn}(\nu_n)=\mathrm{sgn}(\sin(\phi))$. Since this trajectory is an ellipse, non-trivial topology requires
\begin{equation}
\left(\frac{\frac{\bomega}{\Omega}+n}{2\eta_\omega}\right)^2
+\left(\beta-1\right)^2\le1,
\label{eq:boundaries}
\end{equation}
or equivalently, $\left( \frac{\bar{\omega}}{\Omega} + n \right)^2 \le (2\eta_\omega)^2\,\beta(2-\beta).$
This inequality identifies when the local winding number is non-trivial, $\nu_n \neq 0$, namely for $0 < \beta < 2$ and $\eta_\omega,\eta_\kappa > 0$. We stress that the modulation of both frequency and decay channels, via $\eta_\omega$  and $\eta_\kappa$, is essential for obtaining non-trivial topology, as otherwise, Eq. (\ref{eq:boundaries}) traces a horizontal line rather than a closed loop.

Dynamical stability, on the other hand, requires net loss, i.e., $0 < \beta < 1$, see further discussion in Sec.~\ref{sec:Stability_SNR}. The intersection of these conditions determines a topological yet dynamically stable regime. Near the critical point of stability $\beta \simeq 1$, where pump and loss nearly balance, the topological frequency-conversion window, defined by,
\begin{eqnarray}
    \Delta n = 4\eta_\omega \sqrt{\beta(2-\beta)},
\end{eqnarray}
reaches its maximal value $\Delta n = 4\eta_\omega$ ($\Delta\omega = 4\eta_\omega \Omega$).

In Fig.~\ref{fig:Gn_winding}, we observe the main features of topological amplification and frequency conversion. In the topological phase, where $\nu_n(\bar\omega)$ is nontrivial, there is a quasi-zero singular value (i.e., $E_0\approx0$), separated from the bulk singular values by a gap, leading to the approximation $\msG_{nm}(\bar\omega)\approx (\bsb{v}_0)_n (\bsb{u}_0^*)_m/E_0$, where the FGF is dominated by the zero channel. In panels (a–d), we plot the absolute value of the FGF for increasing values of $\eta_P$, such that, as we approach $\beta\rightarrow 1$, the system moves deeper into the topological phase, leading to smaller $E_0$ values. This, in turn, enhances the FGF, giving rise to the well-known phenomenon of topological amplification in the Floquet-Sambe space, and consequently, to frequency conversion. Chiefly, the region where $\nu_n(\bar\omega)$ is nontrivial pinpoints the indices at which $\msG_{nm}$ attains its maximal weight, confirming the predictive power of the local approximation. Moreover, the sign of $\nu_n$ directly indicates up- ($+$) or down- ($-$) frequency conversion.
\section{Generalized Jackiw--Rebbi model}\label{sec::4}
Interestingly, further features of our amplifier, in particular, the input and output bandwidths, controlled by the singular vectors $\bu_0$ and $\bv_0$, can be quantitatively captured by the continuous Jackiw-Rebbi theory, see Fig.~\ref{fig:Gn_winding}(e, f). 

Let us consider, for simplicity and without loss of generality, the doubled Hamiltonian~\eqref{eq:H_k_omegabar_n} at $\bomega=0$ 
\begin{align}
\begin{aligned}
\frac{\tilde{\mcH}_k(0,n)}{\Omega}=&\,(n+2\eta_\omega\sin k)\,\sigma_x
\\
&+\frac{\eta_\kappa}{2}\left(\beta-1-\cos k\right)\sigma_y.    
\end{aligned}
\label{eq:H_cal_omega_bar_0}
\end{align}
Since $\mcH(\bar{\omega})$ has a discrete chiral symmetry, the situation is analogous to the Su–Schrieffer–Heeger model for polyacetylene chains~\cite{Heeger88rmp}, where a topological soliton emerges at the interface between distinct dimerization patterns. Topological-insulator theory predicts that the solitons located at the ellipse boundaries $n_0=\pm|2\eta_\omega\sqrt{\beta(2-\beta)}|$, correspond to zero–singular-value eigenstates of ${\cal H}(\bar{\omega})$, separated from the continuum by a finite gap. This behavior directly mirrors the JR mechanism for a Dirac particle with a spatially varying mass.

\begin{figure}[h]
\centering
\includegraphics[width=.9\linewidth]{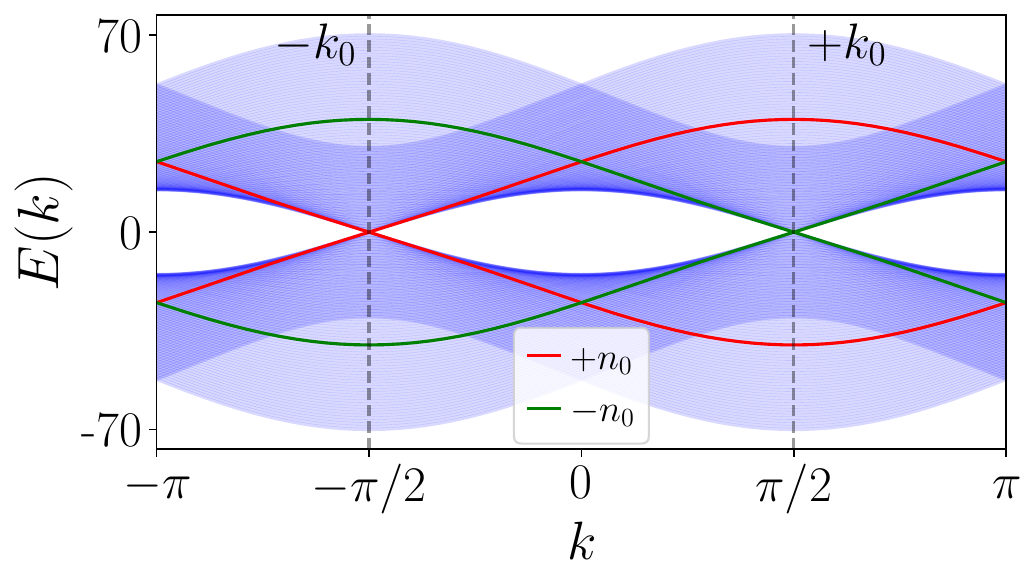}
\caption{Eigenvalues of the doubled matrix $\tilde{\mcH}_k(0,n)$ for different values of $n\in[-50,50]$ and $\beta=1$. The plot shows the presence of two Dirac cones at the values $(+k_0,-n_0)$ and $(+k_0,-n_0)$. The above figure is computed for the following parameters $\eta_P =20 s$,  $\eta_\omega=\eta_{\gamma}/s=\eta_\kappa/s=10$, $\phi=\pi/2$,  $\bomega=0$, and $s=3$. 
}\label{fig:E_k}
\end{figure}

Exploiting the chiral (sublattice) symmetry, the spectrum consists of pairs of eigenvalues $E(k,n)=\pm s_k(n)$, where $s_k(n)\geq0$ denotes the singular value of $\tilde{\mathsf{H}}_k(n)$. The inequivalent minima of $s_k(n)$ occur at the two Dirac points $(+k_0,-n_0)$ and $(-k_0,+n_0)$, see Fig.~\ref{fig:E_k}, where
\begin{align}
\begin{aligned}
k_0&=|\arccos(\beta-1)|, \\
n_0&=|2\eta_\omega\sin k_0|
     =|2\eta_\omega\sqrt{1-(\beta-1)^2}|.
\end{aligned}
\end{align}

 Linearizing around either cone, i.e., $k=\pm k_0+\delta k$ and $n=\mp n_0+\delta n$, yields an effective Dirac Hamiltonian whose spatially varying mass is $m(\delta n)=\delta n$. This model supports Gaussian JR solitons centered at $n=\pm n_0$, with complex widths set by the slope (velocity) of the corresponding cone, which are given by
\begin{align}\label{psiLR}
\begin{aligned}
\psi_L &=
\begin{pmatrix} u(n) \\ 0 \end{pmatrix},
\qquad
\psi_R =
\begin{pmatrix} 0 \\ v(n) \end{pmatrix},
\end{aligned}
\end{align} denoting the left and right solitons localized at $(+k_0,-n_0)$, and $(-k_0,+n_0)$, respectively, 
with
\begin{align}
\begin{aligned}
u(n) &= \mcl{N}_u \,e^{-\frac{(n+n_0)^2}{2\sigma_{\rm r}^2}}
        e^{-i\frac{(n + n_0)^2}{2\sigma_{\rm i}^2}}, \\
v(n) &= \mcl{N}_v \,e^{-\frac{(n-n_0)^2}{2\sigma_{\rm r}^2}}
        e^{-i\frac{(n-n_0)^2}{2\sigma_{\rm i}^2}}.
\end{aligned}
\end{align}
The value of the normalization constants $\mcl{N}_u$ and $\mcl{N}_v$, the widths $\sigma_r$ and $\sigma_i$, and further algebraic details are given in the App.~\ref{App:JRmodel} for the general case. In the ideal topological regime ($\beta=1$), these expressions simplify considerably, i.e., $n_0\to 2\eta_\omega$ while the phase chirp disappears, so that the profile becomes a purely real Gaussian with width $\sigma_{\rm r}=\sqrt{\eta_\kappa/2}$. This reproduces with excellent accuracy the numerically obtained zero–singular-value vectors. In Fig.~\ref{fig:Gn_winding}(e, f), we show non-ideal cases with $\beta\neq 1$ (corresponding to the FGF shown in panels Fig.~\ref{fig:Gn_winding}(c, d)).

\section{Stability and SNR}\label{sec:Stability_SNR}
The topological criterion derived above determines when the doubled-space Hamiltonian supports zero-singular-value solitons, but this alone does not ensure dynamical stability. Stability is governed instead by the long-time behavior of the intracavity photon number, which obeys 
\begin{align}
    \frac{d}{dt} \frac{\langle a^\dagger a \rangle}{\Omega} = \eta_\kappa\left(\beta - 1  \right) \langle a^\dagger a \rangle - \eta_\kappa  \cos(\Omega t) \langle a^\dagger a \rangle + \eta_P\nonumber,
\end{align}
where the last term arises from the noise correlations of the pump channel. This quantity grows without bound whenever there is net energy gain ($\beta>1$). Near the ideal topological regime ($\beta\approx1$), the time required to reach the steady state diverges. See App.~\ref{app:AElimination} for a full comparison between the time-domain dynamics from Eq.~\eqref{eq:adot} and the response obtained from the truncated Green’s-function formalism.

Finally, as a simple operational figure of merit to characterize device performance, we compute the signal-to-noise ratio $\mcl{S}_{\rm out}(t)/\mcl{N}_{\rm out}(t)$ defined by splitting the total number of output photons reflected from port $\gamma$ when a coherent drive $\langle d_{\rm in}(t) \rangle = \alpha_d e^{-i\omega_d t}$ is injected at the same port while all other inputs remain in vacuum, see previous Sec.~\ref{sec:I-O_theory}. Using the input–output relations derived above, the output photon flux can be decomposed as
\begin{align}
\langle d_{\rm out}^\dag(t)\,d_{\rm out}(t)\rangle
=\mcl{S}_{\rm out}(t)+\mcl{N}_{\rm out}(t),
\end{align}
where the coherent contribution (signal) reads
\begin{align}
\mcl{S}_{\rm out}(t)
=\left|\alpha_d\sum_n e^{-i n \Omega t} \msf{R}_{n,n_d}(\bar\omega_d)\right|^2,
\end{align}
and the incoherent contribution (noise) is
\begin{align}
\mcl{N}_{\rm out}(t)
=\frac{1}{2\pi} \sum_{n,m}e^{i(n-m)\Omega t}
\sum_{l}\int_0^{\Omega} d\bar\omega \,
\msf{P}^*_{n l}(\bar\omega)\,\msf{P}_{m l}(\bar\omega).
\end{align}
Here, the signal depends on the reflection matrix $\msf{R}_{nm}$, while the noise originates from the incoherent pump and is fully determined by $\msf{P}_{nm}$, defined in Eqs.~(\ref{eqs:R_nm_P_nm}).

\begin{figure}[h]
\centering
\includegraphics[width=.95\linewidth]{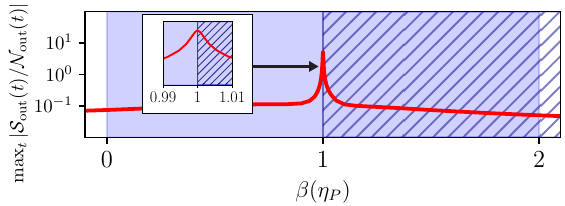}
\caption{Maximum over one modulation period of the time-dependent signal-to-noise ratio defined in the text, computed in the steady state as a function of $\beta(\eta_P)$. The blue (dashed) region identifies the topological (unstable) regime. Parameters for the system are $\eta_\omega=\eta_\kappa=\eta_\gamma=10$, $\phi=\pi/2$, and $\eta_P\in(9,30)$. We set the injected-signal drive frequency to $n_d=-19$  ($\bar{\omega}_d=0$) next to the lower peak of $\bu_0$ (see  Fig.~\ref{fig:Gn_winding}(f)) when $\eta_P\approx 19.5$ ($\beta\approx0.95$), and take the amplitude to be $\alpha_d=1$.}
\label{fig:avg_a_w_SNR_t}
\end{figure}

The resulting SNR exhibits a clear maximum as the system approaches the topological singular regime $\beta \to 1$, corresponding to the boundary between stable (solid) and unstable (dashed) regions in Fig.~\ref{fig:avg_a_w_SNR_t}.
This behavior is explained by the fact that, at $\beta = 1$, the singular vectors $\bsb{u}_0$, $\bsb{v}_0$, are described by Jackiw-Rebbi solitons with a minimal width in frequency space (see App. \ref{App:JRmodel}, Eq. \eqref{widths}). Thus, the efficiency of input signal collection is optimal at this operating point. In a practical setting, since the system is not stable at $\beta = 1$, the optimal operating point would be achieved by approaching the value $\beta \to 1$ until non-linear or saturation effects take over. We note in passing that extending the notion of added noise to Floquet topological amplifiers remains an open problem~\cite{Caves1982,Ramos2021} for future work.

\section{Implementation in cQED}\label{Sec::IMP}
\begin{figure}[h]
\centering
\includegraphics[width=.75\linewidth]{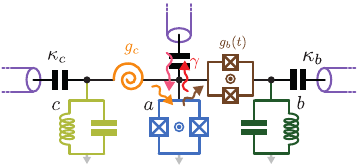}
\caption{Schematic of a proposed cQED setup. The slow mode $a$ couples to two fast-decaying auxiliary modes via flux-pumped Josephson elements. A dc SQUID mediates a time-modulated coupling $g_b(t)$ to the lossy mode $b$ (with decay  $\kappa_b$), while a SNAIL provides a parametric coupling $g_c$ to the pumped mode $c$ ($\kappa_c$). Independent flux lines drive the SQUID at frequency $\Omega$ and pump the SNAIL at a high-frequency  $\omega_p\simeq\omega_a+\omega_c$.}
\label{fig:cQED_implementation}
\end{figure}

The effective model discussed above could potentially be engineered in circuit QED by coupling a slow mode $a$ to fast-decaying auxiliary resonators through flux-pumped Josephson nonlinear elements, see Fig.~\ref{fig:cQED_implementation}. For instance, a SNAIL~\cite{Frattini2018,Sivak2019} provides a clean three-wave-mixing interaction when pumped at $\omega_p \simeq \omega_a + \omega_c$, while a flux-pumped dc SQUID operated in a four-wave-mixing regime implements an analogous interaction when $2\omega_p \simeq \omega_a + \omega_c$~\cite{CastellanosBeltran2008,Planat2019}. When the auxiliary modes have large decay rates, $\kappa_{\rm aux}\gg g,\Omega,\kappa_a,P$, they remain close to vacuum and can be adiabatically eliminated (see App. \ref{app:AElimination}). This yields the effective time-dependent decay rate and incoherent pump used in our model, 
$\kappa(t)=4g_b^2(t)/\kappa_b$ and $P=4 g_c^2/\kappa_c$, where $g_b(t)$ is a modulated beam-splitter coupling and $g_c$ is a parametric pump coupling. These ingredients provide a straightforward and experimentally realistic route to realizing the Floquet topological frequency-conversion and amplification described in this work.

Typical parameter regimes in state-of-the-art superconducting circuits are compatible with the requirements of our model. Microwave resonators in the $4$--$10$~GHz range are routinely used, while auxiliary modes with large linewidths $\kappa_{b,c}/2\pi \sim 100$--$500$~MHz can be engineered to operate in the bad-cavity limit~\cite{Chen2023}. Parametric couplings mediated by flux-pumped Josephson elements typically reach $g/2\pi$ in the few- to few-tens-of-MHz range, with values up to $\sim 80$~MHz demonstrated experimentally~\cite{Lu2023}, yielding characteristic modulated-decay and pump scales of order $1$--$20$~MHz through adiabatic elimination. Choosing a modulation frequency $\Omega/2\pi \sim 5$--$20$~MHz such that the Floquet sidebands remain spectrally resolved relative to the characteristic decay and pump scales gives dimensionless parameters of order unity, with $\eta_\kappa$ setting the amplitude of the modulated decay and $\eta_P=P/\Omega$, $\eta_\gamma=\gamma/\Omega$. Larger values can be accessed either by increasing the parametric coupling strength or by reducing $\Omega$.

The coherent frequency modulation required to realize $\eta_\omega$ can be implemented using SQUID-based tunable resonators, where frequency excursions in the $100$--$500$~MHz range have been demonstrated~\cite{Sandberg2008,Wang2013}. For modulation frequencies in the tens-of-MHz range, this corresponds to $\eta_\omega \sim 5$--$20$. In this regime, the size of the topological window in harmonic space, $\Delta n = 4\eta_\omega \sqrt{\beta(2-\beta)}$, is set primarily by $\eta_\omega$, allowing access to $\Delta n$ of a few tens of sidebands, and up to $\sim 80$ near $\beta\to1$. The parameter $\beta$, which determines both stability and proximity to the topological amplification regime, is also experimentally tunable. In practice, values in the range $0 < \beta \lesssim 1$ can be achieved by balancing the incoherent pump and static loss rates relative to the modulated dissipation, with $\beta \to 1$ enhancing amplification while remaining within the dynamically stable regime.

Finally, we note that decreasing the modulation frequency $\Omega$ increases all dimensionless parameters $\eta_{\alpha}$ for fixed physical rates, thereby improving the continuum (Jackiw--Rebbi) description while keeping constant the topological  frequency-space conversion window $\Delta \omega$.

\section{Conclusions} We have shown that a single driven–dissipative harmonic oscillator can host topological frequency conversion and amplification. Periodic modulation of its frequency and decay generates a non-Hermitian Floquet lattice with asymmetric hopping and a synthetic electric field. A local winding number in doubled space approximately predicts the full phase diagram, including the regime of directional gain and efficient frequency conversion. The associated singular vectors are well captured by Jackiw–Rebbi–type solitons emerging from Dirac cones in synthetic-frequency space. This mechanism is validated numerically and can be implemented, for instance, in circuit QED via modulated dissipation. Our work demonstrates that Floquet-engineered loss is an essential ingredient for enabling topological photonic functionality normally associated with higher-dimensional (physical or synthetic) systems. Looking ahead, this approach may prove useful in quantum-sensing applications and in extensions to richer architectures, ranging from parametrically driven single oscillators to higher-dimensional realizations via multichromatic drives and coupled-oscillator arrays.
\begin{acknowledgments}
A. P.-R. acknowledges support from the Juan de la Cierva fellowship FJC2021-047227-I and from the European Union’s Marie Skłodowska-Curie Actions (MSCA) under grant agreement No. 101204967 (FTMcQED). A. P.-R. and P. G. also acknowledge funding from the Swiss National Science Foundation (Project No.\ CRSII~222812/1). 
This research is part of the Munich Quantum Valley initiative, which is supported by the Bavarian state government with funds from the Hightech Agenda Bayern~Plus. This work is supported by the Spanish projects PID2021-127968NBI00 and PID2024-159152NB-I00 (D. P. and M. C. R.), and by PID2023-146531NA-I00 (T. R. and A. G. L.), financed by MCIN/AEI/10.13039/501100011033 and ERDF/EU. T. R. further acknowledges the Ram\'on y Cajal program RYC2021-032473-I, financed by MCIN/AEI/10.13039/501100011033 and the European Union NextGenerationEU/PRTR.
\end{acknowledgments}

\bibliographystyle{apsrev4-2}
\bibliography{main.bib}

\clearpage
\newpage

\appendix
\section{Adiabatic elimination}
\label{app:AElimination}
Here we derive the time-dependent quantum Langevin equation presented in Eq.~\eqref{eq:adot} of the main text, starting from a microscopic model in which the system mode $a$ is coupled to rapidly decaying auxiliary bosonic $b$ modes. A time-dependent coupling to one auxiliary mode generates an effective modulated decay channel for $a$, while a parametric coupling to a second fast-decaying mode yields an incoherent pump. Both constructions rely on standard adiabatic elimination in the Born-Markov regime.

\subsection{Time-dependent decay channel}

We first consider a single auxiliary mode $b$ that mediates time-dependent decay. The full system is governed by the Hamiltonian
\begin{align}
H = \omega_a a^\dagger a + \omega_b b^\dagger b +  g_b(t) (a^\dagger b + a b^\dagger),
\end{align}
where $g_b(t) = g_{b0} \cos(\Omega t/2)$ is a real-valued time-periodic coupling. The mode $b$ is coupled to a Markovian bath, resulting in fast decay at rate $\kappa_b$, and is taken to be initially in its vacuum state. We also allow mode $a$ to have its own intrinsic loss and (possibly) gain channels, as detailed in the main text.

We move to an interaction picture with respect to the free evolution of the uncoupled modes, defined by
\begin{align}
U_0(t) = \exp[-i(\omega_a a^\dagger a + \omega_b b^\dagger b)t].
\end{align}
In this frame, the interaction Hamiltonian becomes
\begin{align}
\tilde{H}_I(t) = -i g_b(t) \big(a^\dagger b e^{i \Delta_b t} - a b^\dagger e^{-i \Delta_b t}\big),
\end{align}
with $\Delta_b = \omega_b - \omega_a$ the detuning between the two modes. The Heisenberg equations of motion in this interaction picture are
\begin{align}
\dot{a}(t) &= -g_b(t) e^{i \Delta_b t} b(t), \\
\dot{b}(t) &=  g_b(t) e^{-i \Delta_b t} a(t) - \frac{\kappa_b}{2} b(t) +\sqrt{\kappa_b} b_{\text{in}}(t),
\end{align}
where $b_{\text{in}}(t)$ is the input noise operator driving the lossy auxiliary mode, with
$[b_{\text{in}}(t), b_{\text{in}}^\dagger(t')] = \delta(t - t')$.

We assume that the decay rate $\kappa_b$ is large compared to all other relevant energy scales
\begin{align}
\kappa_b \gg g_{b0},\, \Omega,\, \kappa_a,\, P,
\end{align}
so that $b$ can be adiabatically eliminated. Formally integrating the equation for $b(t)$ yields
\begin{align}
b(t) \approx&\, \int_0^\infty d\tau\, e^{-\frac{\kappa_b}{2} \tau}
\left[ g_b(t - \tau) e^{-i \Delta_b (t - \tau)} a(t - \tau)\right.\nonumber\\
&\left.
+ \sqrt{\kappa_b} b_{\text{in}}(t - \tau) \right].
\end{align}
Inserting this expression into the equation for $\dot{a}(t)$, and assuming that $g_b(t)$ and $a(t)$ vary slowly on the timescale $\kappa_b^{-1}$, we use the Markov approximation
$g_b(t - \tau) \approx g_b(t)$ and $a(t - \tau) \approx a(t)$ inside the integral. Evaluating the convolution yields
\begin{align}
\dot{a}(t)
=& -\Gamma_b(t) a(t)
\label{eq:adot_QLE_SM}\\
&- g_b(t) e^{i \Delta_b t} \sqrt{\kappa_b} \int_0^\infty d\tau\, e^{-\frac{\kappa_b}{2} \tau} b_{\text{in}}(t - \tau),\nonumber
\end{align}
with $\Gamma_b(t)=\frac{g_b(t)^2}{\kappa_b/2 - i \Delta_b}$. Writing $\Gamma_b(t) = \kappa(t)/2 + i\delta\omega(t)$, we identify the induced time-dependent decay rate and Lamb shift as
\begin{align}
\kappa(t) &= \frac{4 g_b(t)^2}{\kappa_b^2 + 4 \Delta_b^2}\kappa_b,\qquad
\delta\omega(t) = \frac{4 g_b(t)^2}{\kappa_b^2 + 4 \Delta_b^2}\Delta_b.
\end{align}
On resonance ($\Delta_b = 0$), this simplifies to
\begin{align}
\kappa(t) = \frac{4 g_b(t)^2}{\kappa_b}.
\end{align}
The second term in Eq.~\eqref{eq:adot_QLE_SM} represents an effective input-noise contribution, which can be written as a filtered noise operator. On resonance, it is well-approximated by
\begin{align}
\xi_{\text{aux}}(t)
&= - g_b(t) \sqrt{\kappa_b} \int_0^\infty d\tau\, e^{-\frac{\kappa_b}{2} \tau} b_{\text{in}}(t - \tau)\nonumber\\
&\approx - \frac{2 g_b(t)}{\sqrt{\kappa_b}}\, b_{\text{in}}(t),
\end{align}
consistent with an effective decay channel of rate $\kappa(t)$.

\subsection{Incoherent pump channel}

An incoherent pump channel for $a$ can be engineered in a fully analogous way by coupling $a$ to a second fast-decaying auxiliary mode $c$ via a parametric (two-mode-squeezing) interaction~\cite{Roushan2017,Busnaina2024},
\begin{align}
H_P = \omega_c c^\dagger c + g_c (a^\dagger c^\dagger + a c),
\end{align}
with $c$ coupled to a Markovian bath at rate $\kappa_c$ and initialized in its vacuum state. Working in the appropriate rotating frame and applying the same adiabatic-elimination steps as above (now for $c$), one finds that for
\begin{align}
\kappa_c \gg g_c,\, \kappa_a,\, \kappa_b,\, \Omega,
\end{align}
the mode $c$ induces an effective incoherent pump for $a$ with rate
\begin{align}
P = \frac{4 g_c^2}{\kappa_c},
\end{align}
and an associated noise term $\sqrt{P}\, c_{\text{in}}^\dagger(t)$, where $c_{\text{in}}(t)$ denotes the corresponding input noise operator. This is the standard result for a negative-damping (gain) channel generated via a strongly damped parametric auxiliary mode.

\subsection{Full effective Langevin equation}

Including the intrinsic loss of mode $a$, the engineered time-dependent decay from mode $b$, and the incoherent pump from mode $c$, and restoring the possibly time-dependent system frequency $\omega_0(t)$, we obtain the effective Langevin equation
\begin{align}
\dot{a}(t)
= -i \omega_0(t) a(t)
+ \frac{P - (\gamma + \kappa(t))}{2}\, a(t)
+ \xi_\mathrm{in}(t),
\end{align}
with $\kappa(t) = \frac{4 g_b(t)^2}{\kappa_b}$, where $\gamma$ denotes any additional unmodulated loss channel. The total input noise operator is
\begin{align}
\xi_\mathrm{in}(t)
= -\sqrt{\kappa(t)}\, b_{\text{in}}(t)
  -\sqrt{\gamma}\, d_{\text{in}}(t)
  +\sqrt{P}\, c_{\text{in}}^\dagger(t),
\end{align}
with $d_{\text{in}}$ the input associated with the background loss channel. This Langevin equation coincides with Eq.~\eqref{eq:adot} in the main text and provides a microscopic justification for the effective model with time-dependent decay rate $\kappa(t)$ and incoherent pump $P$. The derivation is valid in the regime $\kappa_{b,c} \gg g_{b0}, g_c, \Omega, \kappa_a$ and under the usual Markov and adiabatic approximations, and it extends straightforwardly to more general slow modulations of $g_b(t)$ and $\omega_0(t)$. 

\begin{figure}
\centering
\includegraphics[width=\linewidth]{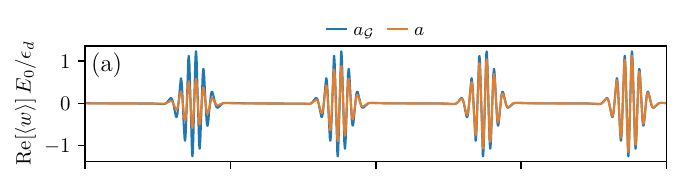}
\includegraphics[width=\linewidth]{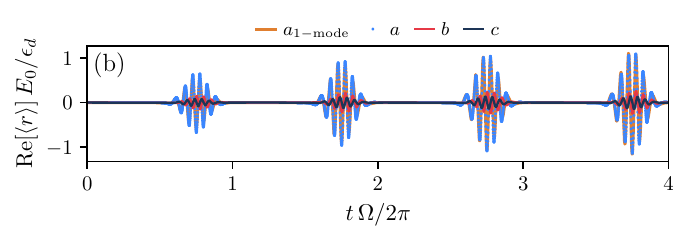}
\caption{(a) Transient dynamics 
for the real part of the averaged fields $\langle w(t)\rangle$, with 
$w\in\{a, a_{\mathcal G}\}$. The curve for the integrated Langevin equation~\eqref{eq:adot}, $w=a$ (orange) approaches the 
steady state response $w=a_{\mathcal G}$ 
(blue), obtained from the truncated Green’s function in 
Eq.~\eqref{eq:abar_wbar_Gnm}. Both quantities are normalized by the drive 
amplitude $\epsilon_d$ and by the singular value $|E_0| \approx 2.4\times10^{-3}$. 
Parameters are $\eta_\omega=\eta_\kappa=\eta_\gamma=10$, $\eta_P=19.8$ 
($\beta=0.98$), and $\phi=\pi/2$. As $\beta\to1$, the system requires 
increasingly long times to reach its steady state. (b) Time-domain comparison between the real part of the amplitudes in the three-mode model ($r=\{a,b,c\}$), and those of the effective (ideal) one-mode model ($r=\{a_{1-\mathrm{mode}}\}$). The parameters used for $a_{1-\mathrm{mode}}$ (and $a$) are the same as in panel (a). The $b$ and $c$ modes are resonant with $a$ and decay at rates $\kappa_b = 200\Omega$ and $\kappa_c = 1200\Omega$, with couplings $g_b(t)$ and $g_c$ chosen to reproduce $\eta_\kappa$ and $\eta_P$.}
\label{fig:time_dynamics}
\end{figure}
As an example, in Fig.~\ref{fig:time_dynamics}(a) we show a comparison between the transient of the real part of the averaged field amplitude $\alpha(t)=\langle a(t)\rangle$ computed by direct integration of the Langevin equation with a coherent input at port $\gamma$, $\langle d_\rmin(t)\rangle =  \alpha_d e^{-i\omega_d t}$ ($\omega_d = \bomega_d + n_d \Omega$) with $\alpha_d=\frac{\epsilon_d}{\sqrt{\gamma}}$, i.e., in frequency space $\langle \bar d_{\rm in,m}(\bar\omega)\rangle = \sqrt{2\pi}\,\alpha_d\,\delta_{m,n_d}\,\delta(\bar\omega-\bar\omega_d)$, and all others in vacuum, with the reconstructed signal obtained from the Green's function formalism (Eq.~\eqref{eq:at_a_n_bomega}), i.e., 
\begin{align}
\langle \bar{a}_n(\bomega) \rangle &= \sqrt{2\pi} \epsilon_d\msf{G}_{n, n_d}(\bomega) \delta(\bomega - \bomega_d).
\label{eq:abar_wbar_Gnm}
\end{align}

In Fig.~\ref{fig:time_dynamics}(b), we show the time-integrated dynamics for the one-mode model and a more realistic three-mode model in cQED, with the two auxiliary modes in the very bad-cavity limit. Naturally, as we approach the topological singular regime ($\beta=1$) the auxiliary degrees of freedom become populated and the effective slow-dynamics deviates from the idealized case.

\section{Solutions to the Generalized JR model}\label{App:JRmodel}
Topological insulator theory predicts the emergence of solitonic modes localized at interfaces separating distinct topological regions. In this Appendix, we present the technical details underlying the derivation of the left- and right-localized Jackiw–Rebbi solitons discussed in Section \ref{sec::4}. In addition, we provide explicit expressions for their normalization constants and spatial widths.

\emph{Left solitonic solution.}
We expand Eq. (\ref{eq:H_cal_omega_bar_0}) around the minimum $(+k_0,-n_0)$ to obtain the left solitonic solution on the $n$-axis by writing $k=k_0+\delta k$, $|\delta k|\ll1$. The Hamiltonian linearized around this point reads
\begin{align}
\frac{\tilde{\mcl{H}}_{\delta k}(0,n)}{\Omega}\approx(n+n_0 + A(\beta)\delta k)\sigma_x
+B(\beta)\delta k\,\sigma_y,
\end{align} 
with the coefficients 
\begin{align}
A(\beta)&\equiv2\eta_\omega\cos k_0=2\eta_\omega(\beta-1),\\
B(\beta)&\equiv\frac{\eta_\kappa}{2}\sin k_0=\frac{\eta_\kappa}{2}\sqrt{\beta(2-\beta)}.
\end{align}
This procedure implements a slowly-varying-envelope (continuum) approximation of the discrete Floquet-Sambe matrix $\bar{\mathsf{H}}_{n m}$ from Eq. \eqref{eq:Hnm_Sambe}. States in the synthetic-frequency basis, $\psi_n$, are represented via a Fourier transform centered around $k_0$:
\begin{align}
    \psi_n=\frac{1}{\sqrt{2\pi}}\int_{\mathbbm{R}} d(\delta k)\,e^{i(k_0+\delta k)n}\psi(\delta k).
\end{align}
Here $\psi(\delta k)$ varies smoothly over the relevant momentum range. To make explicit the slowly-varying-envelope approximation, we factor out from the above expression the fast Bloch oscillations and define a smooth envelope $\varphi(n)$ via
\begin{equation}
\psi_n = e^{ik_0 n} \varphi(n),
\end{equation}
where $\varphi(n)=\frac{1}{\sqrt{2\pi}}\int_{\mathbbm{R}} d(\delta k)\,e^{i\delta k\,n}\psi(\delta k)$ changes only over many synthetic-frequency sites. For notational simplicity, we will simply write $\varphi(n)\equiv \psi(n)$. Under this assumption, neighboring components satisfy
\begin{equation}
\psi_{n\pm1} \simeq e^{\pm ik_0}\big[\psi(n) \pm \partial_n \psi(n)\big].
\end{equation}

In momentum space, this is equivalent to a linear expansion of $\sin(k_0+\delta k)$ and $\cos(k_0+\delta k)$, retaining only first-order terms in $\delta k$. Formally, the mapping $\delta k \mapsto -i\partial_n$ converts the discrete Hamiltonian into a continuous effective Hamiltonian, valid for slowly-varying components. This approximation relies on three assumptions: (i) $|\delta k|\ll 1$, (ii) the envelope varies slowly over several lattice sites, and (iii) higher-order corrections in $\delta k$ or $\delta n_R$ are negligible.

Applying this correspondence, the effective continuous Hamiltonian reads
\begin{align}
\frac{\tilde{\mcl{H}}_{\rm eff}(0,n)}{\Omega}\approx(n + n_0 -iA\,\partial_n)\sigma_x
-iB\,\partial_n\,\sigma_y,
\end{align}
which represents a modified JR Hamiltonian, with two independent velocities $A(\beta)$ and $B(\beta)$, that make the zero mode complex in general. We seek localized (\emph{solitonic}) zero-energy solutions to 
$\tilde{\mcl{H}}_{\rm eff}(0,n) \psi=0$ with $\psi=(u,v)^{\!T}$. The two components satisfy
\begin{align}\label{eq:u_v_right}
\begin{aligned}
(n+n_0) v-(iA+B)\partial_n v&=0,\\
(n+n_0) u-(iA-B)\partial_n u&=0,
\end{aligned}
\end{align}
with Gaussian solutions
\begin{align}
v(n)&=\mcl{N}_v\,e^{\frac{(n+n_0)^2}{2(iA+B)}},\quad
u(n)=\mcl{N}_u\,e^{\frac{(n+n_0)^2}{2(iA-B)}}.
\end{align}
Only $u(n)$ is normalizable, since $B > 0$. Hence,
\begin{align}
\begin{aligned}
\psi_L =
\begin{pmatrix} u(n) \\ 0 \end{pmatrix},
\,\,\,
u(n)=\mcl{N}_u \,e^{-\frac{(n+n_0)^2}{2\sigma_{\rm r}^2}}
      e^{-i\frac{(n + n_0)^2}{2\sigma_{\rm i}^2}}. 
\end{aligned}
\end{align}
The Gaussian widths are given in terms of the velocities by
\begin{equation}
\sigma_{\rm r}^2=\frac{A^2+B^2}{|B|}, \qquad
\sigma_{\rm i}^2=\frac{A^2+B^2}{A},
\end{equation}
and the normalization constant is $\mcl{N}_u=(\pi\sigma_{\rm r}^2)^{-1/4}$.

\emph{Right solitonic solution.}
A similar procedure can be developed by expanding around the minimum $(-k_0,+n_0)$ for the solitonic solution on the right. 
Let $k=-k_0+\delta k$
so that we get a pair of analogous equations to \eqref{eq:u_v_right}, with $(n+n_0)\rightarrow (n-n_0)$ 
after promoting to the continuum, where only $v(n)$ is normalizable since now $B < 0$.
Hence,
\begin{align}
\begin{aligned}
\psi_{R}&=
\begin{pmatrix}0\\v(n)\end{pmatrix},
\,\,\,
v(n)=\mcl{N}_v\,e^{-\frac{(n-n_0)^2}{2\sigma_{\rm r}^2}}
      e^{-i\frac{(n-n_0)^2}{2\sigma_{\rm i}^2}},
\end{aligned}
\end{align}
with the same normalization constant as in the previous case $\mcl{N}_v=\mcl{N}_u$.

The real and imaginary parts of the Gaussian width determine, respectively, the localization and phase curvature of the soliton. Explicitly, using $r=\eta_\kappa/\eta_\omega$,
\begin{align}
\frac{1}{\sigma_{\rm r}^2}
&=\frac{r}{2\eta_\omega}
  \frac{\sqrt{\beta(2-\beta)}}
       {4(\beta-1)^2+\frac{r^2}{4}[\beta(2-\beta)]},\\
\frac{1}{\sigma_{\rm i}^2}
&=\frac{1}{\eta_\omega}
  \frac{2(\beta-1)}
       {4(\beta-1)^2+\frac{r^2}{4}[\beta(2-\beta)]}.
\label{widths}       
\end{align}

At the symmetric point $\beta=1$, the imaginary width contribution vanishes and the Jackiw-Rebbi solitons have their minimum width
\begin{equation}
\sigma_{\rm r}|_{\beta=1}=\sqrt{\eta_\kappa/2},\qquad
1/\sigma_{\rm i}^2|_{\beta=1}=0.
\end{equation}
We can mention in passing that these continuous Gaussian solitonic solutions are self-consistently justified when $\sigma_{r,i}\gg 1$. Interestingly, we observe that the shape of the Gaussian-like singular vectors is increasingly well captured for fixed $\beta$ as  $r\rightarrow \infty$.

Summarizing, topology emerges from the sign change of the effective mass term ($m(n)=n\mp n_0$) in the continuum Dirac description generated by the doubled Hamiltonian $\tilde{\mcl{H}}_{\rm eff}$. This mass inversion produces localized states in the synthetic-frequency dimension, in direct analogy with Jackiw–Rebbi solitons bound to domain walls~\cite{Jackiw1976}. The two symmetry-related minima $(\pm k_0,\mp n_0)$ produce two independent Dirac Hamiltonians with opposite chirality, each hosting a single normalizable zero mode predicted by the index theorem~\cite{Callias1978}. This provides a unified and compact topological explanation for the appearance of protected solitonic states in the synthetic-frequency dimension of the driven amplifier.

\end{document}